\documentclass[aps,prl,twocolumn,preprintnumbers,superscriptaddress,groupedaddress,nofootinbib]{revtex4}
\pdfoutput=1

\usepackage{graphicx}     

\usepackage[bookmarksopen,colorlinks=true,linkcolor=steelblue,
citecolor=orangered,urlcolor=darkred,linktoc=all]{hyperref}

\usepackage{amsmath}
\usepackage{amsfonts}
\usepackage{amssymb}
\usepackage{bm}        
\usepackage{slashed}   

\usepackage{amsopn}
\mathchardef\mhyphen="2D

\usepackage{xcolor}
\definecolor{darkred}{rgb}{0.7, 0., 0.}
\definecolor{orangered}{rgb}{1,0.27,0.}
\definecolor{steelblue}{rgb}{0.275,0.51, 0.706}
\definecolor{forestgreen}{rgb}{0.13,0.55,0.13}
\definecolor{0f5961}{RGB}{15, 89, 97}

\newcommand{\abs}[1]{\left\lvert #1 \right\rvert}
\newcommand{\hyphen}{\,\mathchar`-\mathchar`-\,}

\begin{document}

\preprint{UMN-TH-4219/23,~~FTPI-MINN-23-12}

\title{\Large Muon spin force}

\author{Yohei Ema}
\email{ema00001@umn.edu}
\affiliation{William I. Fine Theoretical Physics Institute, School of Physics and Astronomy,
University of Minnesota, Minneapolis, MN 55455, USA}
\affiliation{School of Physics and Astronomy, University of Minnesota, Minneapolis, MN 55455, USA}
\author{Ting Gao}
\email{gao00212@umn.edu}
\affiliation{School of Physics and Astronomy, University of Minnesota, Minneapolis, MN 55455, USA}
\author{Maxim Pospelov}
\email{pospelov@umn.edu}
\affiliation{William I. Fine Theoretical Physics Institute, School of Physics and Astronomy,
University of Minnesota, Minneapolis, MN 55455, USA}
\affiliation{School of Physics and Astronomy, University of Minnesota, Minneapolis, MN 55455, USA}

\date{\today}
    
\begin{abstract}
Current discrepancy between the measurement and the prediction of the muon anomalous magnetic moment can be resolved in the presence of a long-range force created by ordinary atoms acting on the muon spin via axial-vector and/or pseudoscalar coupling, and requiring a tiny, $\mathcal{O}(10^{-13}\,{\rm eV})$ spin energy splitting between muon state polarized in the vertical direction. We suggest that an extension of the muon spin resonance ($\mu$SR) experiments can provide a definitive test of this class of models. We also derive indirect constraints on the strength of the muon spin force, by considering the muon-loop-induced interactions between nuclear spin and external directions. The limits on the muon spin force extracted from the comparison of $^{199}$Hg/$^{201}$Hg and $^{129}$Xe/$^{131}$Xe spin precession are strong for the pseudoscalar coupling, but  are significantly relaxed for the axial-vector one. These limits suffer from significant model uncertainties, poorly known proton/neutron spin content of these nuclei, and therefore do not exclude the possibility of a muon spin force relevant for the muon $g-2$.

\end{abstract}


\maketitle

\textbf{Introduction}\,---\, 
The next few years will bring an unprecedented increase in intensity of muon sources \cite{Aiba:2021bxe,CGroup:2022tli,Kawamura:2018apy,Thomason:2019fwe}. There are many practical applications of muons such as muon spin resonance ($\mu$SR), muon beams as calibrated sources of neutrinos, and in the far future, the muon collider. At the same time, the studies of muon properties, such as lepton-flavour violation and muon spin dynamics ($g-2$, muon EDM $d_\mu$, muonium spectroscopy), will be brought to a new level of accuracy. 
Currently, the measurement of the muon $g-2$ has put in question the consistency with the Standard Model (SM) theoretical predictions \cite{Aoyama:2020ynm,Borsanyi:2020mff,Muong-2:2021ojo}. The experimental result \cite{Muong-2:2021ojo} for the anomalous magnetic moment is larger than data-driven value predicted in the SM \cite{Aoyama:2020ynm} by about $(3 \hyphen 4) \sigma$, or
\begin{equation}
    \Delta a_{(\mu)} \simeq + 2.5 \times 10^{-9} .
    \label{Da}
\end{equation}
It is tempting to assign this discrepancy to effects of new physics. However, one has to be cognisant of the fact that recent lattice QCD results
\cite{Borsanyi:2020mff,Ce:2022kxy,ExtendedTwistedMass:2022jpw,FermilabLatticeHPQCD:2023jof}, as well as some re-measurement of the di-pion production on $\gamma^* \to 2\pi$ processes \cite{CMD-3:2023alj} point to a larger predicted $a_\mu$, that is perhaps consistent with the experimental value. More analyses as well as independent determinations of the hadronic vacuum polarization are needed to clarify this issue. Nevertheless, it is important to investigate various scenarios under which the muon anomalous magnetic moment can receive a sizeable beyond-SM correction (see {\em e.g.} \cite{Czarnecki:2001pv}).

In this paper, we will examine an alternative suggestion of a muon spin force mediated by a very light force carrier, acting on muon spin {\em at the tree level} and affecting its spin dynamics \cite{Janish:2020knz,Agrawal:2022wjm,Davoudiasl:2022gdg}. Such models almost inevitably operate with sub-eV fields, very weakly coupled to the ordinary atoms and muons. Additional force acting on the muon spin may alter its precession frequency, and therefore can be interpreted as a contribution to $\Delta a_\mu$. Despite smallness of couplings, the long range of a force can lead to a coherent enhancement due to {\em e.g.} all atoms in the Earth enhancing the gradient of the scalar field at Earth's surface where muon spin experiments are performed. The most useful quantity, capable of explaining discrepancy (\ref{Da}) is the muon-force-associated energy splitting between up and down projections of the muon spin,
\begin{equation}
    \Delta E_{(\mu)} = 6\times 10^{-14}\,{\rm eV} 
    \label{DE}
\end{equation}

While significant parts of the available parameter space can be constrained/excluded using the combination of the 5th force constraints and astrophysical limits, it is important to investigate indirect constraints imposed on muon spin force by spin dynamics of ordinary atoms. This has not been performed before, and our paper fills this gap. 
The studies of spins coupled to external directions are well known (see {\em e.g.} \cite{Pospelov:2004fj} and references therein). 
One class of searches tests the coupling of atomic spins to directions created by the external masses, that can be induced by {\em e.g.} $CP$-odd couplings of axions \cite{Moody:1984ba}. An important subset of such experiments searches for the exotic couplings of spin to the vertical direction, approximately parallel to the vector of the Earth's gravitational force, a setup that is most relevant for the muon $g-2$. These experiments can be used, indirectly, to set bounds on exotic interactions of the muon spin. In some sense, the problem is conceptually similar to the problem of muon EDM inducing atomic EDMs, that was shown recently \cite{Ema:2021jds,Ema:2022wxd} to place more stringent bounds on $d_\mu$ compared to direct measurements \cite{Semertzidis:1999kv,Muong-2:2008ebm}. In this paper, we investigate the constraints imposed by the atomic experiments on exotic muon interactions. We find that the constraints are generally stronger by about an order of magnitude than the level needed to explain the discrepancy (\ref{Da}). However, we also note that the calculations are beset by significant nuclear uncertainties and near-accidental cancellations, that prevent us in deriving stronger limits. 

The second goal of this work is to point out that new dedicated experiments with muons, similar to $\mu$SR in low/medium magnetic field, can be used to detect/constrain the anomalous muon spin rotation. We show that already existing muon facilities can be used to test values of the spin splitting (\ref{DE}) relevant for the $g-2$ discrepancy explanation.

\textbf{Muon spin force and $g-2$}\,---\,
We assume the existence of the muon spin force mediated by a very light scalar field:
\begin{align}
{\cal L}_\mathrm{eff} =...&+\frac{1}{2}(\partial_\nu \phi)^2 - \frac{1}{2}m_\phi^2\phi^2  -g_S \phi (\bar nn + \bar pp) \nonumber\\
&+\frac{g^{(\mu)}_A}{2 m_\mu} \partial^\alpha \phi \times \bar \mu \gamma_\alpha\gamma_5 \mu
-{g_P^{(\mu)}}  \phi \times \bar \mu i\gamma_5 \mu. 
\label{Leff}
\end{align}
Notice that as far as the muon spin dynamics is concerned, and in the lowest order in $\phi$, both axial $g^{(\mu)}_A$ and pseudoscalar $g^{(\mu)}_P$ couplings give equivalent effects. 
However, they are not equivalent at the loop level, as they would predict {\em e.g.} different coefficients for the $F_{\alpha\beta}\tilde F^{\alpha\beta}$ coupling below the muon mass, leading eventually to different strength of the coupling to the atomic spin \cite{Pospelov:2008jk,Flambaum:2009mz}. 
In particular, $g^{(\mu)}_A$ does not generate couplings to on-shell photons at low energy, and therefore is immune to searches of photon birefringence along the lines of recent proposal \cite{Fedderke:2023dwj}.

If we take one nucleon and one muon separated by distance $r$, this will result in a potential acting on spin, 
\begin{equation}
H^{(\mu)}_{s\hyphen m} = 
\frac{g_S (g^{(\mu)}_A+g^{(\mu)}_P)}{4\pi\times 2 m_\mu} (\boldsymbol{\sigma}^{(\mu)}\cdot \boldsymbol{\nabla}) \frac{\exp(-m_\phi r)}{r}
\end{equation}
Here ``$s\hyphen m$" stands for ``spin-mass", and superscript $(\mu)$ refers to the type of particles. 
$g_S$ is limited by gravitational fifth-force searches; $g_A$ is limited by stellar energy loss (\cite{Agrawal:2022wjm} and references therein). 

First, we would like to establish a tentative size of the $g_S g^{(\mu)}_P$ suggested by the results of the muon $g-2$ experiment, if one interprets the current discrepancy (\ref{Da}) as a real effect. 
Suppose that the deviation in the muon $g-2$ experiment is caused by the muon spin interacting 
with the mostly vertical gradient created by $\phi$ which should be the case for all ranges of $\phi$ force, $m_\phi^{-1}$, 
exceeding several meters. We are going to combine all constants within one quantity that has dimension of energy, and introduce a vertical unit vector $\mathbf{n} = (0,0,1)$ aligned with $\boldsymbol{\nabla} \phi$. 
Then the additional term in the Hamiltonian for the muon spin propagating in the 
$g-2$ ring can be written as 
\begin{equation}
H = \Delta E_{(\mu)} (\mathbf{s}\cdot\mathbf{n}) = \Delta E_{(\mu)}\, s_z,
\label{Heff}
\end{equation}
with $s_z = \sigma_z/2$.
Using Heisenberg equations of motion, 
we get in the rest frame of muons, $
\frac{d s_i}{d\tau} = i[H,s_i]= \epsilon_{ijk} n_j s_k \times \Delta E_{(\mu)} .$ Here $\tau$ is the proper time. 
This implies that spin precesses around the vertical direction with the cyclic frequency $\Delta \omega = \Delta E_{(\mu)}/\hbar$. The equation needs to be  transformed to the lab frame, using $\gamma \simeq 29.3$,\footnote{
	This magic momentum is used at the BNL/FNAL experiments.
	The J-PARC experiment plans to use a smaller momentum~\cite{Abe:2019thb},
	which results in a larger effect of the muon spin force~\cite{Davoudiasl:2022gdg}.
} 
$d t = \gamma d\tau$, reducing the precession frequency by a gamma factor, 
$\gamma^{-1}  \Delta E^{(\mu)}/\hbar$ (notice that the axial-vector $\boldsymbol{\nabla} \phi$
and the tensor $\mathbf{B}$, $\mathbf{E}$ transform differently under the boost). The size of the effect that would account for the observed $\Delta a_{(\mu)}$ discrepancy (\ref{Da}) can be determined from the following 
equation: 
\begin{equation}
\Delta E_{(\mu)} = (\hbar \omega_{a}) \times \gamma \times (\Delta a_{(\mu)}/a_{(\mu)}),
\end{equation}
where $\omega_a$ is the main precession frequency measured by the $g-2$ experiment and $a_{(\mu)} \simeq \alpha/(2\pi)$. Substituting 
relevant quantities we arrive to the required value of the spin coupling given in Eq. (\ref{DE}).

\textbf{Indirect constraints}\,---\, 
The two main experiments that set indirect constraints on $\Delta E_{(\mu)}$ compare spin precession around the vertical directions of mercury isotopes, $^{199}$Hg and $^{201}$Hg \cite{PhysRevLett.68.135}, and xenon isotopes, $^{129}$Xe and $^{131}$Xe \cite{Zhang:2023qmu} that all have closed electron shells and nonzero nuclear magnetic moments. Defining the interaction of $i$'th isotope via nuclear spin operator $\mathbf{I}$, nuclear magneton $\mu_N$ and the $g$-factor $g_i$ as 
\begin{equation}
    H_i = -\mu_N g_i ({\bf I \cdot B})+\Delta E_{i} (\mathbf{I}\cdot\mathbf{n}),
\end{equation}
one can formulate the constraints on exotic spin interactions as follows:
\begin{align}
	\label{Hg}
   | \Delta E_{\rm Hg} |
   &= \left | \Delta E_{201} - \Delta E_{199}\,\rho_{\mathrm{Hg}} \right| < 3.0\times 10^{-21}\,{\rm eV},\\
   | \Delta E_{\rm Xe} |
   &= \left | \Delta E_{129} - \Delta E_{131}\,\rho_{\mathrm{Xe}} \right| < 1.7\times 10^{-22}\,{\rm eV},
   \label{Xe}
\end{align}
with $\rho_{\mathrm{Hg}} = g_{201}/g_{199} = -0.369139$ and $\rho_{\mathrm{Xe}} = g_{129}/g_{131} = -3.37337$ the gyromagnetic ratios.
Note that the sensitivity to exotic spin force disappears if the $\Delta E$ for two atomic species scale exactly as $g$ factors. More generally, the status of atomic magnetometry relevant for these types of experiments was reviewed recently in \cite{Terrano:2021zyh}. Other notable limits (sensitive to electron and proton couplings to Earth gravity/vertical direction) are reported in \cite{JacksonKimball:2017zky}. Note that the electron-spin-oriented torsion-pendulum experiment \cite{Heckel:2008hw} also sets the constraint on the electron spin coupling to Earth's gravity, which corresponds to $\Delta E_{(e)} < 6\times 10^{-18}$\,eV in our notations.

\begin{figure}[t]
	\centering
 	\includegraphics[width=0.75\linewidth]{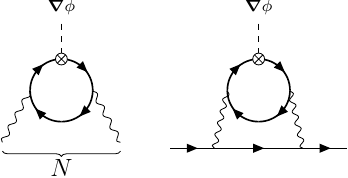}
	\caption{\small The Feynman diagrams corresponding to one- and two-loop effective Lagrangians.
	The thick line corresponds to the muon, and the cross dot indicates the axial-vector/pseudoscalar coupling insertion.
        The electromagnetic fields of the one-loop effective Lagrangian are sourced by the nucleus.}
	\label{fig:Feynman}
\end{figure}

At the next step we construct one- and two-loop effective Lagrangians, integrating out the muon loop,
whose diagrams are shown in Fig.~\ref{fig:Feynman}. 
We treat parameters $g_A^{(\mu)}$ and $g_P^{(\mu)}$ in (\ref{Leff}) as an input, and calculate the {\em induced} couplings
to the external electromagnetic and $\phi$ fields at low energy,
\begin{align}
	{\cal L}_{eff} &=  -\frac{e^2}{16\pi^2 m_\mu}\left(g_P^{(\mu)} \phi F_{\mu\nu}\tilde F_{\mu\nu} 
	+ \frac{g_A^{(\mu)}\partial_\nu \phi}{3 m_\mu^2}\tilde F_{\alpha\nu}\partial_\mu F_{\alpha\mu}\right)
	\nonumber\\
	&= \frac{e^2 \boldsymbol{\nabla} \phi}{4\pi^2 m_\mu} \cdot \left(g_P^{(\mu)} \mathbf{B}A_0 
	+ \frac{g_A^{(\mu)}}{12m_\mu^2} \mathbf{B}\,\nabla^2 A_0\right).
	\label{EB}
\end{align}
In this expression we retained the leading order terms in $1/m_\mu$ expansion for each coupling, and neglected higher derivative terms acting on $\phi$, $\nabla_i\nabla_j \phi \to0$. The $g_A^{(\mu)}$ coupling does not generate corrections to the propagation of photons. In the second line of (\ref{EB}) we used time independence of all background fields, and introduced electrostatic potential $A_0$ via $\mathbf{E} = -\boldsymbol{\nabla} A_0$. 

As discussed in \cite{Flambaum:2009mz}, in the presence of interaction (\ref{EB}), the internal nuclear electromagnetic fields will lead to the coupling of the nuclear spin to $\boldsymbol{\nabla} \phi$. In order to calculate observables $\Delta E_{\rm Hg,Xe}$ in (\ref{Hg}) and (\ref{Xe}) induced by (\ref{EB}), one needs detailed knowledge of the EM field distribution inside these nuclei, which is relatively simple for $A_0(\mathbf{r})$ as it is dominated by collective effects, $A_0\propto Z$, 
and quite difficult for $\mathbf{B}(\mathbf{r})$.

In our evaluation, we employ the following phenomenological approach. We take the magnetic field to be the sum of the one created by the valence nucleon (which happens to be a neutron for all these nuclei) and core polarization that is needed to describe observed magnetic moments for these nuclei. In other words, 
\begin{equation}
   \mathbf{B}(\mathbf{r}) = \mathbf{B}_n(\mathbf{r})+\mathbf{B}_{\rm core}
   (\mathbf{r})
   \xrightarrow{r\gg R_N}
   \frac{1}{4\pi}\boldsymbol{\nabla}\times(
   \boldsymbol{\mu}\times \boldsymbol{\nabla})\frac{1}{r}, 
   \label{bc}
\end{equation}
where $\boldsymbol{\mu}$ is the observed magnetic moment of a given nucleus. Boundary condition (\ref{bc}) does not allow full reconstruction of the magnetic field profiles. In order to achieve that, we solve for the valence neutron wave functions numerically, using one particle Schr\"odinger equation in a collective nuclear potential, and treating neutrons in $^{129,131}$Xe and $^{199,201}$Hg nuclei as $s_{1/2}$, $d_{3/2}$, $p_{1/2}$, $p_{3/2}$ orbitals, 
as described in \cite{Dmitriev:2003sc}. 
This determines magnetic spin density, the spin current, and ultimately $\mathbf{B}_n(\mathbf{r})$ everywhere, including inside the nucleus. For $\mathbf{B}_{\rm core}$, we make a phenomenological model that it corresponds to a sphere with constant magnetic polarization, following an empirical fact of near constant nuclear density inside the nuclear radius $R_N$,
and fix its normalization by requiring that the sum~\eqref{bc} reproduces the observed nuclear magnetic moments 
($\mathbf{B}_n$ alone does not quite reproduce the observed values except for ${}^{199}$Hg).

Following the steps outlined above we calculate $\Delta E_i$'s entering (\ref{Hg}) and (\ref{Xe}). The details of the calculation are provided in the Appendix. Parametrically, we have the following size of the nuclear spin coupling to the vertical direction,
\begin{align}
	\label{gP}
 \Delta E_{i} &= \Delta E_{(\mu)} \times b_i \frac{Z\alpha^2(\mu_n/e)}{4\pi R_{Ni}}~~{\rm for}~g_P^{(\mu)},\\
    \Delta E_{i} &= \Delta E_{(\mu)} \times c_i \frac{Z\alpha^2(\mu_n/e)}{4\pi R_{Ni}^3 m_\mu^2}~~{\rm for}~g_A^{(\mu)},
    \label{gA}
\end{align}
where $\mu_n$ is the neutron magnetic moment, and $R_{Ni}$ are corresponding nuclear radii
(related to the charge radii $r_c$ as $R_N = \sqrt{5/3}\,r_c$). Constants $b_i,c_i$ encode numerical coefficients that arise from the numerical evaluation of the matrix elements needed to convert (\ref{EB}) to $\Delta E_{i}$. Substituting numerical results, we get the following prediction for the observables: 
\begin{align}
\label{gP1loop}
    \abs{\frac{\Delta E_{\rm Hg}}{\Delta E_{(\mu)}}} &= 1\times 10^{-6};~ 
	\abs{\frac{\Delta E_{\rm Xe}}{\Delta E_{(\mu)}}} = 3 \times 10^{-6} ~{\rm for}~g_P^{(\mu)},\\
	\abs{\frac{\Delta E_{\rm Hg}}{\Delta E_{(\mu)}}} &= 2\times 10^{-8};~ 
	\abs{\frac{\Delta E_{\rm Xe}}{\Delta E_{(\mu)}}}  = 4\times 10^{-9} ~{\rm for}~g_A^{(\mu)}. 
    \label{gA1loop}	
\end{align}
Coupling $g_P^{(\mu)}$ is constrained stronger than $g_A^{(\mu)}$, and if pseudoscalar coupling dominates the muon spin force, the muon $g-2$ anomaly speculation is ruled out by the indirect effect, as the constraint from Xe implies $\abs{\Delta E_{(\mu)}} < 6\times 10^{-17}\,\mathrm{eV}$, which is 
three orders of magnitude stronger than (\ref{DE}). This is not the case for $g_A^{(\mu)}$ coupling. 
When interpreted as an upper bound, the results for $g_A^{(\mu)}$ formally imply $|\Delta E_{(\mu)}| < 4\times 10^{-14}$\,eV
from Xe which is somewhat stronger than the re-interpretation of the muon $g-2$ discrepancy (\ref{DE}). Notice, however, that the results are extremely sensitive to the assumptions about the nuclear magnetic fields, so that the coefficients in (\ref{gA1loop}) are much smaller than the parametric estimate in (\ref{gA}) and result from strong $\mathcal{O}(1\%)$ numerical cancellations. 
Therefore, the re-interpretation of (\ref{Hg}) and (\ref{Xe}) constraints into a firm limit on $\Delta E_{(\mu)}$ is problematic.

Now we turn to the calculation of the two-loop induced spin couplings. 
The two-loop effects induced by the $g_P^{(\mu)}$ coupling is UV-finite (if the Higgs boson contributions are neglected). The calculation of the induced electron coupling is well defined (see {\em e.g.} \cite{Flambaum:2019ejc,Ema:2021jds}),
\begin{equation}
   g_P^{(e)} =  g_P^{(\mu)} \times \frac{3\alpha^2m_e}{2\pi^2m_\mu}\times \log(m_\mu/m_e),
\end{equation}
but is small and does not lead to competitive bounds. The $g_P^{(\mu)}$-induced coupling to nucleons is also suppressed. The calculation cannot be done with free quarks, and has to be performed at a nucleon level by combining virtual E1 and M1 transitions. The most important nucleon, on account of \cite{PhysRevLett.68.135,Zhang:2023qmu}, is the neutron that has a very suppressed coupling to the electric field. Parametrically, the two-loop effect will then be suppressed by 
$\alpha^2 r_n^2 m_\rho^2 m_\mu (\mu_n/e) $, where $r_n^2 m_\rho^2$
is the product of the neutron charge radius and $\rho$-mass {\em i.e.} the approximate hadronic scale of convergence for electromagnetic loops. This combination is small, and subdominant to (\ref{gP}). Therefore, we conclude that the main effect of $g_P^{(\mu)}$ is given by one loop, in Eq. (\ref{gP1loop}).  

In contrast, the two-loop effects induced by $g_A^{(\mu)}$ can be important for two reasons. 
It is well known that $g_A$ couplings will give a logarithmic divergence at two loops, requiring the introduction of an ultraviolet cutoff ({\em i.e.} the scale of UV completion) $\Lambda_{\rm UV}$ which can partially compensate for loop smallness. The second reason is that the one-loop effect in (\ref{gA}) are suppressed by $(R_N^3m_\mu^2m_p)^{-1}\sim 3\times 10^{-3}$ while two-loop effects are not. The two-loop result at the elementary particle level is given by
\begin{align}
	{\cal L}_\mathrm{eff} &= \sum_{i=u,d,s,e}\frac{g_A^{(i)}}{2m_\mu} 
	\partial_\alpha \phi \times \bar \psi_i \gamma^\alpha\gamma_5 \psi, \nonumber\\
	g_A^{(i)}&= -g_A^{(\mu)} \times \frac34 \left(\frac{\alpha}{\pi}\right)^2 
	Q_i^2 \log(\Lambda_{\rm UV}^2/\Lambda_{\rm IR}^2),
\label{gAi}	
\end{align}
where $Q_e=-1,\,Q_u = 2/3$ etc. $\Lambda_{\rm IR}$ should be identified with the hadronic scale for quarks, $\Lambda_{\rm IR}\simeq m_p$, while for the electrons the integration can be extended further down to $\Lambda_{\rm IR}\simeq m_\mu$. 
UV cutoff depends on details of UV completion, but should not parametrically exceed $m_\mu/g_A^{(\mu)} $. It is another fundamental uncertainty of the low-energy model (\ref{Leff}). 

 Further model dependence is exacerbated by the fact that other couplings and other particles can contribute to $g_A^{(i)}$ at the same loop level. For example, above the EW scale, $W^+W^-$ and $\gamma Z$ loops will contribute to the running of the couplings, and one needs a proper $SU(2)\times U(1)$ completion of (\ref{Leff}) to extend the treatment above $m_W$. Thus, for example, a low energy {\em vector} coupling of muons, $g^{(\mu)}_V/(2 m_\mu)\partial^\alpha \phi \times \bar \mu \gamma_\alpha \mu$ will contribute  into $g_A^{(i)}$ due to running  above the electroweak scale.

At the same time, it is easy to see that $\Lambda_{\rm UV}$ can be much lower than the electroweak scale. For example $\mu\hyphen\tau$ symmetry \cite{Karshenboim:2014tka} can be employed to have a full cancellation of muon and tau loops above $\Lambda_{\rm UV} = m_\tau$. Given that the couplings of $\phi$ to tau leptons are unconstrained, this choice is perhaps the most conservative. In that case, the renormalization group logarithm, as far as the quark couplings are considered, is given by $\log(m_\tau^2/m_p^2) \simeq 1$.

Now we would like to estimate the sensitivity of Xe and Hg experiments to $g_A^{(i)}$ and ultimately to $g_A^{(\mu)}$. Using the neutron beta decay measurements and lattice calculations \cite{FlavourLatticeAveragingGroupFLAG:2021npn}, one can translate quark axial couplings to those of the nucleons:
\begin{eqnarray}
    g_A^{(p)}=  0.777g_A^{(u)}-0.438g_A^{(d)} -0.053g_A^{(s)};\nonumber\\ g_A^{(n)}= -0.438g_A^{(u)}+0.777g_A^{(d)}-0.053g_A^{(s)},
\label{gn}
\end{eqnarray}
where we have used (2+1+1)-light fermion results \cite{FlavourLatticeAveragingGroupFLAG:2021npn}. 

At the next step, we need to combine neutron and proton spins into the total spins of Xe and Hg isotopes. The isotopes of Xe and Hg all have unpaired neutrons over the closed shells, but the accuracy of such a model in predicting observable $\mu_i$ is rather poor for some isotopes (for example, for $^{201}$Hg). Moreover, the method where a difference between observed magnetic moment and the shell model prediction (Schmidt model) of $\mu_i$ is attributed to the proton spin magnetism (see {\em e.g.} \cite{Flambaum:2009mz}) seems to over-predict the nuclear proton spin content \cite{JacksonKimball:2014vsz}. 

In what follows, we consider the Xenon results (\ref{Xe}) using the nuclear calculations of the spin content from~\cite{Klos:2013rwa}, as well as two other nuclear models quoted in the same paper. Combining the dominant neutron, and subdominant proton contributions using additional inputs from (\ref{gAi}) and (\ref{gn}), we arrive at
\begin{equation}
\Delta E_{\rm Xe} = (3 \hyphen 7) \times 10^{-8}\times \Delta E_{(\mu)} \times \frac{\log(\Lambda_{\rm UV}^2/\Lambda_{\rm IR}^2)}{\log(m_\tau^2/m_p^2)}.
\end{equation}
 The uncertainty in this coefficient comes mostly from the variation of the nucleon spin content of Xe isotopes, from model to model. Still, it appears to be dominant over one loop result of (\ref{gA1loop}) by a factor of a few, implying a limit on muon spin force, 
 $\vert \Delta E_{(\mu)}\vert < 6\times 10^{-15}$\,eV. 

We conclude that the results of the Xe and Hg experiments set indirect constraints on $\Delta E_{(\mu)}$ about one order of magnitude stronger than the value needed to reconcile the predicted and measured values of the muon $g-2$, Eq.~(\ref{DE}). However, we note that there are sources of multiple cancellations at the hadronic and nuclear level, as well as strong sensitivity to UV completions of such models that negates the utility of such bounds. For example, a $\mathcal{O}(10\%)$ accidental cancellation of different sources inside $g_A^{(n)}$ would completely negate the utility of these bounds to challenge the $g-2$ suggested value. Therefore, one should check for the presence/absence of $\Delta E_{(\mu)}$ directly, using the high-intensity muon sources.

\textbf{Anomalous spin precession of stopped muons}\,---\,The effect of an exotic spin force (\ref{Heff}) induces spin-precession of the stopped $\mu^+$ particles. If magnetic field is completely shielded, the muonic spin force will introduce the rotation angle during one muon lifetime
\begin{eqnarray}
    \Delta \psi = \frac{\Delta E_{(\mu)}t}{\hbar}
    ~\Longrightarrow ~ \Delta \psi
    (t=\tau_\mu) = 2\times 10^{-4},
\end{eqnarray}
where $\tau_\mu= 2.2\,{\rm\mu s}$ is muon lifetime.
The minimum number of stopped muons required to detect such an angle, assuming large efficiency of the positron detection, is given by $N_{(\mu)} > 1/(\Delta \psi)^2 \sim 10^8$. With modern $\mu$SR beam lines capable of delivering over $10^5$ muons per second, this appears to be entirely feasible. 

On the other hand, the {\em equivalent} magnetic field that would introduce same strength rotation is $B^{eq} =\Delta E_{(\mu)}/\mu_{\mu}\simeq 1.1\,{\rm mGs}$,  where $\mu_{\mu}$ is the muon intrinsic magnetic moment. The smallest DC magnetic fields detected with $\mu$SR techniques are about 0.1\,Gs \cite{RevModPhys.69.1119}. It is then clear that the muon spin force effect needs to be measured as the shift of muon spin precession frequency at low/moderate values of the magnetic field. In the past, studies of the muon magnetic moment stopped in liquid helium in the field of 66\,Gs have brought a relative precision of $\mathcal{O}(10^{-4})$ \cite{PhysRevLett.33.572}. One order of magnitude improvement in accuracy would reveal or exclude $\Delta E_{(\mu)}$.

A suggested $\mu$SR setup for an experiment should involve the following ingredients:
\begin{itemize} 
\item Uniform vertical magnetic field in the place where muons are stopped is required. A high degree of uniformity across the stopping target should be possible to achieve. The optimal magnetic field values that will afford precision measurement of the muon spin precession frequency $\omega_{(\mu)}$ will correspond to $\Delta \psi \propto \mathcal{O}(1)$. This corresponds to the magnetic fields in excess of 10\,Gs, and we suggest the range of 5-to-50 Gs, and a possibility of reversing polarity.

\item A co-magnetometer species that monitors the magnetic field and its stability is needed. As in the muon $g-2$ experiment, one can use proton (water-based) or $^3{\rm He}$ co-magnetometers \cite{Aoyama:2020ynm,PhysRevLett.124.223001}. 

\item The highest accuracy results can be achieved in the gaseous stopping target, that would also afford atomic co-magnetometry. This will minimize the medium feedback on the field and on the muon spin. Low densities of such targets necessitate low-momentum of incoming muons. The upcoming high-intensity muon beams at  PSI \cite{Aiba:2021bxe} will provide $10^5$/sec muon beams with $\sim 4$\,MeV momentum (reduced relative to a usual 29\,MeV for surface muons) that will stop within 1\,cm inside a gaseous target. 

\item For a continuous muon beam, $10^5$/sec intensity appears limiting to minimize the muon pile-up, while the pulsed muon beam with a sub-$\mu$s pulse duration can afford higher muon intensities. 

\end{itemize}

If both muon and co-magnetometer spin precession frequencies are measured, their ratio $r_{\mu/p} = \omega_{(\mu)}/\omega_{(p)}$ should not depend on the value of the magnetic field, in the absence of the spin force. Therefore, performing the measurement at several different values of the magnetic field should isolate the signal of interest. The following quantity, 
\begin{equation}
  -B \frac{dr_{\mu/p}}{dB} = \frac{\Delta E_{(\mu)}}{2\mu_p B} = 3.4\times 10^{-4}\times \frac{\Delta E_{(\mu)}}{6\times 10^{-14}\,{\rm eV}}\times \frac{{\rm 10\,Gs}}{B},
\end{equation}
provides a figure of merit that needs to be achieved in this suggested experiment. Here we used proton spin precession as reference.
The effect is obviously enhanced at small values of the magnetic field, and reverses sign with the reversal of $B$ direction. The effect also changes sign for $\mu^-$, which can be tested experimentally by studying negative muon spin precession in muonic atoms (if challenges due to lower fluxes of $\mu^-$ and muon depolarization during the cascade can be overcome).

While $\mu$SR setting appears to be the most promising place for a muon spin force search, it is not the only way of achieving this goal. Other possibilities include precision studies of muonium hyperfine structure in low/medium vertically directed magnetic field \cite{Kanda:2020mmc}.

\textbf{Conclusions}\,---\,
In this \emph{Letter}, we have investigated terrestrial probes of the muon spin force
mediated by a light $CP$-violating scalar field.
In the first part, we derived indirect constraints on the muon spin force from Hg and Xe isotope experiments,
with the muon appearing only inside quantum loops.
We found that the pseudoscalar coupling explaining the muon $g-2$ anomaly is well-excluded by these experiments.
Although the axial-vector coupling is also excluded by an order of magnitude with our modeling of nuclear physics, 
this constraint is subject to huge nuclear uncertainty, casting doubts on its robustness. 
This motivates us to suggest, in the second part,
a new $\mu$SR-type experiment to look for the muon spin force directly.
Based on our estimation, an experiment testing the muon spin force required for the $g-2$ anomaly 
is entirely feasible within the current technology.
Our study thus offers an interesting experimental opportunity for future muon precision physics facilities.

\vspace{3.5mm}
\begin{acknowledgments}
\textbf{Acknowledgments}\,---\,
Y.E. and M.P. are supported in part by U.S. Department of Energy Grant No.
desc0011842. We would like to acknowledge helpful communications with Drs. D. Budker and M. Romalis. 
Y.E. would like to thank the Aspen Center for Physics (supported by National Science Foundation grant PHY-2210452) 
where part of this work was done.
M.P. would like to thank Perimeter Institute for hospitality during the completion of this project. T.G. would like to thank the Institute for Nuclear Theory and the Perimeter Institute for the visits and discussions about this project.
The Feynman diagrams in this paper are drawn with \texttt{TikZ-Feynman}~\cite{Ellis:2016jkw}.
\end{acknowledgments}

\bibliographystyle{utphys}
\bibliography{MuSpin_main}

\clearpage
\appendix
\onecolumngrid

\renewcommand{\thesection}{S\arabic{section}}
\renewcommand{\theequation}{S\arabic{equation}}
\renewcommand{\thefigure}{S\arabic{figure}}
\renewcommand{\thetable}{S\arabic{table}}
\renewcommand{\thepage}{S\arabic{page}}
\setcounter{equation}{0}
\setcounter{figure}{0}
\setcounter{table}{0}
\setcounter{page}{1}

\begin{center}
\textbf{\Large Supplemental Material
}
\end{center}
In this supplemental material, we provide details on our one-loop, nuclear magnetic field, and two-loop calculation.
\section{Coupling to the external electromagnetic field at one-loop}
Here we start from the muon spin force \eqref{Leff} and derive the one-loop effective Lagrangian \eqref{EB}. 

We first calculate the effect induced by $ g_A^{(\mu)} $. Following \cite{Novikov:1983gd}, the one-loop effective action in the external electromagnetic field can be written as 
\begin{equation}
    \begin{aligned}
    \label{Seff}
        S_{A,\mathrm{eff}}&=-\frac{g_A^{(\mu)}}{2m_\mu}\partial_\nu \phi\times \text{Tr}\left[\frac{i}{i\slashed{D}-m_\mu}\gamma^\nu \gamma^5\right],
    \end{aligned}
\end{equation}
where the trace is over the spinor indices and the spacetime, and $D_\alpha$ is the covariant derivative, which is given by
\begin{equation}
    iD_\alpha=i\partial_\alpha+eA_\alpha,
\end{equation}
and $e$ is the charge of the positron. Using $(i\slashed{D})(i\slashed{D})=(iD)^2+\frac{1}{2}e (\sigma F)$, where $(\sigma F)=\sigma_{\alpha\beta}F^{\alpha\beta}$, we expand \eqref{Seff} in terms of the external field. Up to the order of our interest, we get
\begin{equation}
    \label{SeffinT}
    S_{A,eff}=-i\frac{g_A^{(\mu)}}{2m_\mu}\partial_\nu \phi \times 
    \left(T_0^\nu-\frac{e}{2}T_1^\nu+\frac{e^2}{4}T_2^\nu\right),
\end{equation}
where
\begin{align}
    &T_0^\nu=\text{Tr}\left[\frac{1}{(iD)^2-m_\mu^2}(i\slashed{D})\gamma^\nu\gamma^5\right],\\
    &T_1^\nu=\text{Tr}\left[\frac{1}{(iD)^2-m_\mu^2}(\sigma F)\frac{1}{(iD)^2-m_\mu^2}(i\slashed{D})\gamma^\nu\gamma^5\right],\\
    &T_2^\nu=\text{Tr}\left[\frac{1}{(iD)^2-m_\mu^2}(\sigma F)\frac{1}{(iD)^2-m_\mu^2}(\sigma F)\frac{1}{(iD)^2-m_\mu^2}(i\slashed{D})\gamma^\nu\gamma^5\right].
\end{align}
In the expression above, we omitted terms with $m_\mu$ in the numerator because they vanish after taking the trace over spinor indices. $T_0^\nu$ also vanishes due to the trace property of gamma matrices.

To evaluate $T_1^\nu$, we use the fact that the quantity $\text{Tr}\left[f(F)\frac{1}{(iD)^2-m_\mu^2}\right]$ does not depend on the momentum $iD_\alpha$, where $f(F)$ is an arbitrary function of the electromagnetic field. By shifting the momentum by $q_\alpha$ and focusing on terms linear in $q_\alpha$, we get
\begin{equation}
    \text{Tr}\left[f(F)\frac{1}{(iD)^2-m^2}(iD)_\alpha\frac{1}{(iD)^2-m^2}\right]=0.
\end{equation}
Therefore,
\begin{equation}
    T_1^\nu=\text{Tr}\left[F^{\alpha\beta} \frac{1}{(iD)^2-m_\mu^2} (iD)_\gamma \frac{1}{(iD)^2-m_\mu^2}\right]\text{tr}\left[\sigma_{\alpha\beta}\gamma^\gamma\gamma^\nu\gamma^5\right]=0.
\end{equation}
To evaluate $T_2^\nu$ up to $(DF)F$ order, the following identities turn out to be useful
\begin{equation}
    \begin{aligned}
    \label{derivatives}
        \left[(iD)^2,\sigma F\right]&=\left[iD_\alpha,\left[iD^\alpha,\sigma F\right]\right]+2\left[iD_\alpha,\sigma F\right]iD^\alpha=2iD_\alpha(\sigma F)iD^\alpha+\cdots,\\
	   \left[(iD)^2,iD_\nu\right]&=\left[iD_\alpha,\left[iD^\alpha,iD_\nu\right]\right]+2\left[iD_\alpha,iD_\nu\right]iD^\alpha=-eD^\alpha F_{\alpha\nu}+2ieF_{\alpha\nu}iD^\alpha=0+\cdots,
    \end{aligned}
\end{equation}
where terms that introduce extra electromagnetic fields or more than one derivative on the electromagnetic field are not of our interest. Now $T_2^\nu$ can be written as
\begin{equation}
\label{T2}
    \begin{aligned}
        T_2^\nu&=\text{Tr}\left[\frac{1}{(iD)^2-m^2}(\sigma F)\frac{1}{(iD)^2-m^2}(\sigma F)\frac{1}{(iD)^2-m^2}(iD)^\alpha\gamma_\alpha\gamma^\nu\gamma^5\right]\\
        &=\text{Tr}\left[\frac{1}{\left((iD)^2-m^2\right)^3}(\sigma F)(\sigma F)(iD)^\alpha\gamma_\alpha\gamma^\nu\gamma^5\right]\\
        &\quad +\text{Tr}\left[\frac{1}{\left((iD)^2-m^2\right)^4}\left(2\left[(iD)^2,\sigma F\right](\sigma F)+(\sigma F)\left[(iD)^2,\sigma F\right]\right)(iD)^\alpha\gamma_\alpha\gamma^\nu\gamma^5\right]+\cdots.\\
    \end{aligned}
\end{equation}
By looking at terms linear in $ q $ in $ \text{Tr}\left[f(F)\frac{1}{\left((iD-q)^2-m^2\right)^2}\right] $, one can show that the first term in \eqref{T2} vanishes up to a total derivative. The second term already has enough powers of $F$ and $DF$, so everything commutes. Using \eqref{derivatives}, we arrive at the following result 
\begin{equation}
    \begin{aligned}
        T_2^\nu&=\frac{ig^{\rho\zeta}}{2}\text{Tr}\left[\frac{(iD)^2}{((iD)^2-m^2)^4}\right]\left[\left(D_\rho F^{\alpha\beta}F^{\gamma\delta}+D_\rho(F^{\alpha\beta} F^{\gamma\delta})\right)\right]\text{tr}\left[\sigma_{\alpha\beta}\sigma_{\gamma\delta}\gamma_\zeta\gamma^\nu\gamma^5\right]=-\frac{i}{6\pi^2 m_\mu^2}\int d^4 x{\tilde{F}_\alpha}^{~\nu}D_\beta F^{\alpha\beta},\\
    \end{aligned}
\end{equation}
where we performed integration by parts and neglected terms proportional to $\partial_\alpha\partial_\beta\phi$. Putting it back to \eqref{SeffinT}, we get
\begin{equation}
    \label{SAeff}
    S_{A,\mathrm{eff}}=-\frac{e^2g_A^{(\mu)} \partial^\nu \phi}{48\pi^2 m_\mu^3}\int d^4 x \tilde{F}_{\alpha\nu}D_\beta F^{\alpha\beta}.
\end{equation}
We have checked that we reproduce the same result by the direct Feynman diagrammatic computation
of the triangle diagrams.
The effect induced by $g_P^{(\mu)}$ is obtained directly from the anomaly equation: 
\begin{equation}
    \partial_\nu(\bar{\mu}\gamma^\nu\gamma^5\mu)-2m_\mu\bar{\mu}i\gamma^5\mu=-\frac{e^2}{8\pi^2}F_{\alpha\beta}\tilde{F}^{\alpha\beta},
\end{equation}
multiplying both sides by $\frac{g_P^{(\mu)\phi}}{2m_\mu}$ and performing integration by parts leads to
\begin{equation}
    -g_P^{(\mu)}\phi\bar{\mu}i\gamma^5\mu=-\frac{e^2g_P^{(\mu)}}{16\pi^2m_\mu}\phi F_{\alpha\beta}\tilde{F}^{\alpha\beta}+\frac{g_P^{(\mu)}\partial_\nu\phi}{2m_\mu}\bar{\mu}\gamma^\nu\gamma^5\mu.
\end{equation}
As is shown in \eqref{SAeff}, the contribution from $\bar{\mu}\gamma^\nu\gamma^5\mu$ is suppressed by the muon mass, so at leading order, only the first term on the right-hand side contributes, so we get
\begin{equation}
    \mathcal{L}_{P,\mathrm{eff}}=-\frac{e^2g_P^{(\mu)}}{16\pi^2m_\mu}\phi F_{\alpha\beta}\tilde{F}^{\alpha\beta}.
\end{equation}
Again, we have checked that we obtain the same result by computing the triangle diagrams directly.

\section{Nuclear magnetic field}
Here we provide details on our model of the magnetic field in \eqref{bc} which leads to \eqref{gP1loop} and \eqref{gA1loop}. The non-relativistic eigenfunction of a spin 1/2 particle in spherical coordinates can be written as \cite{Berestetskii:1982qgu}
\begin{equation}
    \psi(r,\theta,\varphi)=R_{nl}(r)\Omega_{jlm}(\theta,\varphi),
\end{equation}
where
\begin{equation}
    \begin{aligned}
        \Omega_{l+1/2,l,m}=\left(\begin{array}{c}
				\sqrt{\frac{j+m}{2j}}Y_{l,m-1/2}\\
				\sqrt{\frac{j-m}{2j}}Y_{l,m+1/2}
			\end{array}\right),\quad 
                \Omega_{l-1/2,l,m}=\left(\begin{array}{c}
				-\sqrt{\frac{j-m+1}{2j+2}}Y_{l,m-1/2}\\
				\sqrt{\frac{j+m+1}{2j+2}}Y_{l,m+1/2}
			\end{array}\right),
    \end{aligned}   
\end{equation}
$ Y_{l,m} $ are spherical harmonics, and the radial function $ R_{nl}(r) $ is determined by
\begin{equation}
    \frac{d^2(rR_{nl})}{dr^2}+2m(E_n-U)rR_{nl}-\frac{l(l+1)}{r^2}rR_{nl}=0.
\end{equation}
They are normalized by
\begin{equation}
    \int d\Omega \left(\Omega_{j'l'm'}^\dagger\Omega_{jlm}\right)=\delta_{jj'}\delta_{ll'}\delta_{mm'},\quad \int dr r^2 R_{nl}R_{n'l}=\delta_{nn'}.
\end{equation}
The magnetic field generated by the magnetic moment of the valance neutron is given by \cite{Landau:1991wop}
\begin{equation}
    \boldsymbol{B}_n(\boldsymbol{r})=-\frac{\mu_n}{4\pi}\int d^3r'\left[\boldsymbol{\nabla}'\times\left(\psi^\dagger(\boldsymbol{r}')\boldsymbol{\sigma}\psi(\boldsymbol{r}')\right)\right]\times\boldsymbol{\nabla}\frac{1}{|\boldsymbol{r-r'}|},
\end{equation}
where $\mu_n$ is the magnetic moment of the neutron and $\boldsymbol{\sigma}$ are Pauli matrices.

While it is possible to compute the magnetic field directly for any given eigenstate, a much simpler expression can be obtained when we average out the angular dependence, this is valid as long as the electric potential $A_0$ is independent of the angle. The angular averaged magnetic field is calculated to be
\begin{equation}
\begin{aligned}
    \label{Bave1}
    \boldsymbol{\mathcal{B}}_n(r)&=\frac{1}{4\pi}\int d\Omega\boldsymbol{B}_n(\boldsymbol{r})\\
    &=\frac{\mu_n}{(4\pi)^2}\int d^3r'\left[\boldsymbol{\nabla}'\times\left(\psi^\dagger(\boldsymbol{r}')\boldsymbol{\sigma}\psi(\boldsymbol{r}')\right)\right]\times\int d\Omega\frac{\boldsymbol{r-r'}}{|\boldsymbol{r-r'}|^3}\\
    &=-\frac{\mu_n}{4\pi}\int d^3r'\left[\boldsymbol{\nabla}'\times\left(\psi^\dagger(\boldsymbol{r}')\boldsymbol{\sigma}\psi(\boldsymbol{r}')\right)\right]\times\hat{\boldsymbol{r}}'\frac{\Theta(r'-r)}{r'^2},
\end{aligned}
\end{equation}
where $\Theta(x)$ is the Heaviside step function. The integration over $d\Omega'$ involves only spherical harmonics and the result can be parameterized as
\begin{equation}
    \int d\Omega'\left[\boldsymbol{\nabla}'\times\left(\psi^\dagger(\boldsymbol{r}')\boldsymbol{\sigma}\psi(\boldsymbol{r}')\right)\right]\times\hat{\boldsymbol{r}}'=\left(c_1\frac{R_{nl}^2}{r'}+c_2\frac{\text{d}R_{nl}^2}{\text{d} r'}\right)\hat{\boldsymbol{I}},
\end{equation}
where $\hat{\boldsymbol{I}}$ is the direction of the nuclear spin, and $c_1$ and $c_2$ are constants which depends on the angular quantum numbers. For the cases of our interest, they are
\begin{equation}
    (c_1,c_2)_{{}^{129}\text{Xe}}=\left(0, \frac{2}{3}\right),
    \quad
    (c_1,c_2)_{{}^{131}\text{Xe}}=\left(-\frac{6}{5}, -\frac{4}{5}\right),
    \quad
    (c_1,c_2)_{^{199}\text{Hg}}=\left(-\frac{4}{3},-\frac{2}{3}\right),
    \quad
    (c_1,c_2)_{^{201}\text{Hg}}=\left(\frac{2}{5},\frac{4}{5}\right).
\end{equation}
Now \eqref{Bave1} simplifies to
\begin{equation}
    \boldsymbol{\mathcal{B}}_n(r)=-\frac{\mu_n \hat{\boldsymbol{I}}}{4\pi}\left[c_1\left(\int_r^\infty dr'\frac{R_{nl}^2(r')}{r'}\right)-c_2R_{nl}^2(r)\right]
\end{equation}
The magnetic moment of the nucleus generated by the valance neutron is obtained from the relation $\int d^3r \boldsymbol{B}_n(\boldsymbol{r})=\frac{2}{3}\mu_N\hat{\boldsymbol{I}}$, it is related to $c_1$ and $c_2$ by
\begin{equation}
    \mu_N=-\frac{\mu_n}{2}(c_1-3c_2).
\end{equation}
The difference between the valance neutron contribution to nucleus magnetic moment $\mu_N$ and 
the experimental value $\mu_{N,\text{exp}}$ (taken \emph{e.g.} from~\cite{Stone:2005rzh}) 
is modeled by a uniformly magnetic polarized core, averaging over all directions, we get
\begin{equation}
    \boldsymbol{\mathcal{B}}_{\text{core}}(r)=\frac{\mu_{N,\text{exp}}-\mu_N}{2\pi R_N^3}\hat{\boldsymbol{I}}\Theta(R_N-r).
\end{equation}

To get numerical results, we numerically solve the radial wave function taking $U$ to be a Woods-Saxon potential using the parameters in \cite{Dmitriev:2003sc}. The numbers in \eqref{gP1loop} and \eqref{gA1loop} are then obtained from \eqref{EB} using
\begin{equation}
    H=-\int d^3 r \mathcal{L}_\mathrm{eff}.
\end{equation}

\section{Coupling to light fermions at two-loop}

\begin{figure}[t]
    \centering
    \includegraphics[width=0.65\linewidth]{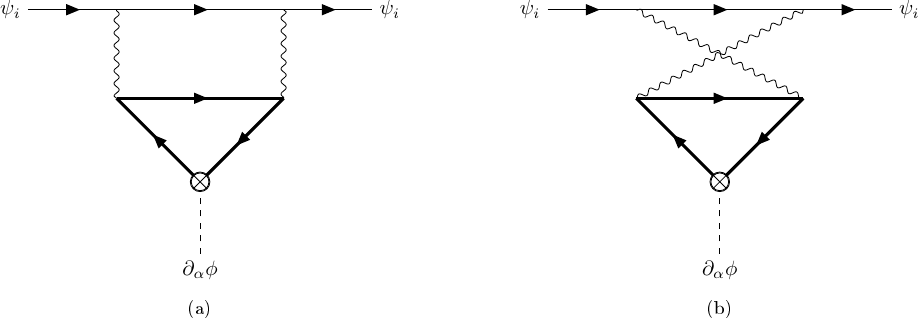}
    \caption{Two-loop diagrams leading to Eq.~\eqref{gAi}.
    The upper fermion line represents the light fermion and the lower fermion loop is formed by the muon.}
    \label{2_loop}
\end{figure}

Here we discuss the computation that leads to \eqref{gAi}. The contribution comes from the two diagrams shown in Fig.~\ref{2_loop}. We are interested in the UV divergent part of the diagram, so all the external momentum as well as the mass of the light fermion are set to 0. Following the procedure described in \cite{Ema:2022wxd}, we write the amplitude as
\begin{equation}
    \mathcal{M}=\frac{g_A^{(\mu)}\partial_\alpha\phi}{2m_\mu}\bar{\psi}_i\mathcal{M}^\alpha \psi_i,
\end{equation}
where $\mathcal{M}_\alpha$ can be factored into a scalar integral and a term describing its Lorentz and spinor structure:
\begin{equation}
    \mathcal{M}^\alpha=\frac{1}{24d(d-1)(d-2)(d-3)}\text{tr}\left[\mathcal{M}_\kappa\epsilon^{\kappa\lambda\nu\xi}\gamma_\lambda\gamma_\nu\gamma_\xi\right]\times\epsilon^{\alpha\beta\gamma\delta}\gamma_\beta\gamma_\gamma\gamma_\delta.
\end{equation}
After the integration-by-parts reduction, the two-loop integral reduces to two one-loop integrals, which are computed using the standard Euclidean space technique. The divergent part of our result is given by
\begin{equation}
    \mathcal{M}=-\frac{g_A^{(\mu)}}{2m_\mu}\frac{3}{4}\left(\frac{\alpha}{\pi}\right)^2Q_i^2\ln\left(\frac{\Lambda_{\text{UV}}^2}{\Lambda_\mathrm{IR}^2}\right)\partial_\alpha\phi\times \bar{\psi}_i\gamma^\alpha\gamma^5\psi_i.
\end{equation}
In the case of the electron, the IR scale is given by $\Lambda_\mathrm{IR} = m_\mu$.
In the case of the up and down quarks, the QCD scale kicks in before the energy scale reaches $m_\mu$,
below which the free quark picture is no longer valid.
Therefore we may take the IR scale as the hadronic scale, \emph{e.g.}, $\Lambda_\mathrm{IR} = m_p$.

\end{document}